\pdfoutput=1
\documentclass[11pt]{article}

\usepackage[T1]{fontenc}
\usepackage[utf8]{inputenc}
\usepackage{mathptmx}                 
\usepackage[scaled=0.92]{helvet}      
\usepackage{courier}                  

\usepackage[letterpaper,margin=1.15in]{geometry}
\usepackage{microtype}
\usepackage{framed}                   
\usepackage{enumitem}
\usepackage[hyphens]{url}
\usepackage{xcolor}
\definecolor{inklink}{rgb}{0.13,0.18,0.33}
\usepackage[colorlinks=true,breaklinks=true,
            linkcolor=inklink,
            urlcolor=inklink,
            citecolor=inklink]{hyperref}

\usepackage{titlesec}
\titleformat{\section}{\normalfont\sffamily\large\bfseries}{\thesection}{0.6em}{}
\titleformat{\paragraph}[runin]{\normalfont\bfseries}{}{0em}{}[.]
\titlespacing*{\paragraph}{0pt}{1.0em}{0.6em}

\hypersetup{
  pdftitle={The Substrate Collapse: AI Code Generation Invalidates Authorship-Based Knowledge Metrics},
  pdfauthor={Brett Wheeler}
}

\title{The Substrate Collapse:\\[2pt]
  AI Code Generation Invalidates Authorship-Based Knowledge Metrics}
\author{Brett Wheeler\\[2pt]
  Independent Researcher\\
  \href{mailto:bwheeler@brettwheeler.com}{bwheeler@brettwheeler.com}\\
  ORCID: \href{https://orcid.org/0009-0000-2744-489X}{0009-0000-2744-489X}}
\date{June 2026}

\begin{document}
\maketitle

\begin{center}
\itshape\small
Why the truck factor, Degree-of-Authorship, and the degree-of-knowledge family no
longer measure what they were built to measure --- and why the replacement cannot be
a reweighting of them.
\end{center}

\begin{abstract}
\noindent
Software engineering has long inferred where a system's knowledge resides from who
authored its code. The truck factor, the Degree-of-Authorship metric, and the
degree-of-knowledge model all rest on a single inference --- that authoring a region
of code is evidence of understanding it --- and for most of software's history that
inference was a workable proxy, because for code to enter a repository a human
generally had to write it, and writing it usually forced at least a transient
understanding. This paper argues that AI code generation now severs that inference at
its root, and that the consequence is not the degradation of the authorship-based
metrics but their invalidation as a class. When an agent generates a module and a
human merges it, the version-control record still attributes authorship, but the
attribution no longer licenses any conclusion about comprehension: the same footprint
is now compatible with full understanding, partial understanding, or none at all. The
metric still returns a number. The number measures a substrate that has come uncoupled
from the quantity it was used to estimate. I show that this collapse is corroborated by
the field's own measurement failures --- a flagship productivity experiment whose
follow-up could no longer reliably size the effect once developers grew unwilling to
work without the tools it studied, a large-scale telemetry instrument that admits in
its own report that it cannot distinguish rework-waste from repair, and a corpus study
in which generation is competent at surface defects and fails precisely where program
understanding lives. The methodological corollary is the load-bearing claim: the
instrument the comprehension-debt era needs cannot be built by refining the
knowledge-concentration metrics, because no function of an authorship footprint
recovers an inference the footprint no longer supports. The replacement must be
grounded in evidence of comprehension rather than evidence of authorship. I state a
falsifiable prediction that discriminates the two --- that systems with a healthy
authorship-derived truck factor but low comprehension-measured retention will suffer
incident-resolution failures the authorship metric does not predict --- and argue that
building the comprehension-grounded instrument at the scale of a system and a team is
the field's open measurement problem, deliberately left open here.
\end{abstract}

\section{A relocated constraint, and the instrument the field reaches for}

The marginal cost of producing source code has collapsed, and the collapse is no
longer an estimate. By 2026 the authorship of production code by AI tools is directly
observable: a majority of developers use an AI coding tool weekly, most teams cross a
high threshold of weekly active use, and the rate at which AI-generated code is
accepted into codebases has climbed steeply over two years [Faros 2026; Wang 2025].
And the generation is no longer merely voluminous; it is increasingly autonomous ---
agent-mode tooling plans, executes, and corrects its own work, with a research line now
folding reflection into the agent's reasoning loop so that it monitors and repairs its
own output before review [Wang 2026b]. The obvious reading of this collapse --- drawn
explicitly by the spec-driven-development vendors and implicitly by most practitioners
--- is that software has become cheap to produce and therefore cheap to rebuild
[GitHub 2025]. The reading this paper assumes, rather than argues, is the opposite: the
collapse does not make software easy, it relocates the binding constraint, moving it
from the activity that got cheap --- production --- to the activities that did not.
Those are verification, and beneath verification, comprehension: holding a working
understanding of what a system does and why. The relocation has been named, in quick
succession and under at least three labels, as \emph{comprehension debt} [Gorman 2025;
Osmani 2026; Ahmad 2026], \emph{cognitive debt} [Storey 2026; Thoughtworks 2026], and
\emph{epistemic debt} [Sankaranarayanan 2026].

A field that has just named a constraint reaches for an instrument to measure it, and
the instrument it reaches for is always the nearest one already in hand. For the
question \emph{who still understands this system, and how concentrated is that
understanding,} the nearest instrument is the knowledge-concentration literature: the
truck factor [Avelino et al. 2016], the Degree-of-Authorship metric, the
degree-of-knowledge model [Fritz et al. 2014], and the turnover-induced
knowledge-loss studies built on them [Rigby et al. 2016]. These are mature, automated,
and already on the wall. A team that wants a number for \emph{how exposed are we if the
wrong person leaves} computes a truck factor today, and the computation runs without
complaint.

This paper argues that this reflex is a mistake, and a particular kind of mistake worth
naming precisely. The knowledge-concentration metrics are not strained by AI generation
in a way that any calibration could repair. They are invalidated as a class, by the
same development that makes the measurement they promise urgent. The instrument still
reads. It reads the wrong substrate. What follows states the single inference the
family rests on (\S2), shows AI generation severing it and argues the severance is
class-wide and unrecoverable by reweighting (\S3), reads three of the field's own
recent measurement strains as corroboration of it, not proof (\S4), says what survives
the collapse while declining to build it (\S5), and stakes a prediction that exposes the
whole claim to refutation (\S6).

The contribution is negative, and the negativity isn't a hedge --- it's the point. This
paper proposes no replacement metric, specifies no instrument, and discloses no
mechanism for measuring comprehension. It establishes that an entire family of working,
trusted metrics has had its foundation removed, that the removal cannot be undone from
inside the family, and that the field is therefore measuring knowledge concentration
with an instrument that has come loose from the thing it reports. Saying why the old
instrument is dead is the precondition for building the right one. It is not the
building.

\section{The one inference the metric family rests on}

The knowledge-concentration tradition measures where understanding sits in a
development organization, and it is the closest existing analogue to what the
comprehension-debt literature now wants read. Its instruments are well specified and,
on their home ground, good. The truck factor --- the minimal number of developers whose
departure would incapacitate a project --- was given a rigorous automated estimator by
Avelino et al. [2016], who found that roughly two-thirds of 133 popular repositories had
a truck factor of two or fewer. The machinery underneath is a model of authorship: the
Degree-of-Authorship metric, and the more elaborate degree-of-knowledge model of Fritz
et al. [2014], infer a developer's familiarity with a region of code from the footprint
they have left on it in version control --- commits, lines authored, edits, interaction
history. Rigby et al. [2016] turned the same machinery on the question of how much
knowledge an organization loses when its developers turn over.

Strip the family to its load-bearing element and one inference remains, holding all of
it up:

\begin{leftbar}
\noindent\textbf{The authorship inference.} A developer's footprint on a region of code
is evidence of their understanding of that region. \emph{Authored $\Rightarrow$
understands.}
\end{leftbar}

Every metric in the family is this inference wearing different clothes. The truck factor
is a count of the people the inference lets you call knowledgeable about each part of
the system, aggregated into a number that reads as robustness. Degree-of-knowledge is
the inference made quantitative and per-developer. Turnover knowledge-loss is the
inference applied to the people walking out the door. Remove it and nothing in the
family computes a quantity that means anything; the version-control statistics survive,
but they no longer stand for understanding, and standing for understanding was the whole
job.

The metrics are recent --- the truck factor and its kin are a decade or so old --- but
the inference they formalize is as old as the practice of programming, and for most of
that history it held. Its soundness had a mechanical cause rather than a fortunate one:
for a region of code to enter a repository, a human generally had to write it, and
writing a non-trivial region usually forced its author to build --- at least for the
duration --- a theory of what it should do, in Naur's [1985] sense: a living
understanding of how the system behaves, why it is shaped as it is, and how it can be
changed without coming apart. The footprint was the fossil that theory left in the
record. It was never a clean fossil, and the field knew it --- a developer can author
and forget, can paste a snippet they never understood, can merge boilerplate by rote ---
and Fritz et al. [2014], whose model this paper cites, found that a developer's
knowledge showed up in interaction data the authorship record never captured. The proxy
leaked. But it leaked at the edges and tracked across the body, because the act that
produced the footprint was, in the main, the act that produced the understanding. The
metric never measured understanding directly. It measured a trace that understanding
usually left.

\section{The collapse}

AI generation does not invent the gap between a footprint and an understanding ---
\S2's proxy already leaked. What it does is destroy the mechanism that held the leak to
the edges, and it destroys it in the ordinary case, not the corner.

Take the event that now happens a thousand times a day: an agent generates a module, and
a human merges it. The merge may follow genuine review, in which the human reconstructed
a theory of the code before accepting it. It may follow a skim. It may follow no review
at all --- the pattern the comprehension-debt literature describes, in which generated
changes land directly through agent-mode tooling at rates that have risen sharply
[Faros 2026]. In every one of these cases the version-control record attributes
authorship --- to the human committer, to an agent identity, to both --- and in every
one of them the attribution says \emph{nothing} about comprehension. The human's
footprint is now compatible with full understanding, with partial understanding, and
with none, and the footprint itself cannot tell you which. The act that produced it ---
merging --- is no longer the act that produces understanding, because the writing that
once forced understanding was done by a system that does not retain it as theory a human
could be asked to defend.

What \S2 called a leak at the edges now runs through the whole surface. The signal and
the quantity it proxied have come apart. This is the substrate collapse, and its
consequence is exact. A truck factor computed over an AI-generated codebase still
returns a number, and the number is meaningless --- not approximately, but in the
precise sense that the thing it measures, the distribution of authorship footprint, has
stopped being correlated with the thing it was used to estimate, the distribution of
retained theory. The instrument still reads. It just reads the wrong substrate. This is
not the wrongness of a miscalibrated gauge, which reports a biased version of the right
quantity and can be set right with a constant. It is the wrongness of a thermometer in a
world where fever and temperature have come uncoupled: the reading is perfect, as
faithful and as precise as it ever was, and it has simply ceased to be evidence of the
illness anyone picks up a thermometer to find. The instrument did not fail. The
correlation it stood on did.

Two consequences follow, and the second is the one practitioners will fight.

\paragraph{The invalidation is class-wide} It does not strike one metric and spare the
rest, because the metrics do not fail independently --- they fail at the single joint
they share. Any metric whose definition routes through an authorship footprint to reach
a conclusion about understanding inherits the collapse whole. The truck factor,
Degree-of-Authorship, degree-of-knowledge, and the turnover-loss models built on them go
down together, not as a coincidence but as a structural fact about what they are made of.

\paragraph{The collapse cannot be repaired by reweighting the footprint} Here is the
natural defense, and it is the right move to try: weight commits by review depth, by
interaction time, by edit complexity; subtract agent-attributed lines; discount the
merges that closed too fast to have been read. Each of these is a function of the
authorship footprint, and not one of them recovers the inference --- for a reason that
is structural, not a matter of finding better weights. After the collapse, \emph{every}
value of the footprint is compatible with full, partial, or zero comprehension, because
the human's relationship to AI-generated code is no longer constrained by the footprint
at all. A developer may hold a deep theory of a module they never touched in version
control, having read it and reasoned it through; a developer may carry a heavy footprint
on a module they merged without understanding a line. No reweighting of a quantity that
has decoupled from its target can re-couple it, because the information needed to recover
the inference --- does this person actually understand this code --- is simply not
present in the authorship record, in any weighting, the generative act having cut the
link that once put it there. Recovering it means measuring something else: comprehension,
which lives on a different substrate (\S5).

This is why the verdict is invalidation and not recalibration, and why the
methodological corollary is the real weight of the paper. The instrument the
comprehension-debt era needs cannot be built by refining the knowledge-concentration
metrics, because refining them only sharpens a measurement of a quantity that has
stopped mattering. You cannot tune your way back to a correlation that no longer exists.

\section{Measurement under the new regime}

The argument of \S3 stands on its own; this section does not prove it but corroborates
it. If the authorship substrate has come loose from understanding, we would expect the
field's wider measurement apparatus to be straining too --- failing not at
comprehension, which it never measured, but at coarser quantities, in ways consistent
with instruments reading a process that has moved out from under them. Three recent
results show that strain. None is new data, and none directly measures the
authorship-comprehension link; each is a reading of public evidence, offered as
corroboration and labelled as such.

\paragraph{The perception gap, observed under controlled conditions} METR's randomized
controlled trial of early-2025 tools found experienced developers on familiar codebases
measurably slower with AI assistance --- about 19\% on the central estimate --- while
believing themselves roughly 20\% faster, the slowdown traced largely to the overhead of
reviewing and integrating generated output [METR 2025]. That inverted self-assessment is
the durable result, and the one that serves this argument: developers were wrong
\emph{in direction} about their own productivity, under controlled conditions --- the
false-confidence signature the comprehension-debt literature predicts, caught on an
instrument built to catch it. METR's larger follow-up on 2026 tools could not extract a
reliable estimate, and said why with unusual candor: developers increasingly declined to
participate because they were unwilling to work without AI --- the primary cause, and one
that biases the measured speedup downward --- compounded by a pay-rate reduction that
introduced a selection effect of its own and by unreliable time-on-task measurement for
the developers running several agents at once [METR 2026]. METR's own read is that the
tools likely do speed developers up in early 2026, but that this design can no longer
size the effect. The lesson worth drawing is the narrow one, narrower than the episode
invites: a clean controlled estimate of even AI's effect on raw speed has become hard to
obtain, in part because the adoption under study has begun to erode the control condition
itself. If the simplest quantity now resists clean measurement, the subtler quantity this
paper is concerned with --- which no instrument is tracking at all --- will not be easier.

\paragraph{The instrument that confessed} Large-scale industry telemetry across thousands
of developers shows the signature of a relocated bottleneck with unusual clarity: more
tasks completed and many more pull requests merged, alongside steep rises in
pull-request size, bugs per developer, and time in review [Faros 2025; Faros 2026], and
--- in the more recent cross-section --- a tripling of incidents per pull request and a
sharp rise in the share of pull requests merged with no review at all [Faros 2026]. One
figure does double duty as a lesson in metrology. Code churn --- lines deleted relative
to lines added in recently merged code --- rose dramatically; the report itself lists
three candidate explanations for the rise --- rework of generated code that failed in
practice, long-deferred refactoring finally being tackled, and engineers returning sooner
to code they were never satisfied with --- and concedes, in its own pages, that the
metric as measured cannot resolve among them. The industry's largest telemetry
instrument can see that the codebase is convulsing; it cannot see whether the convulsion
is waste or repair --- which is to say it cannot read the one distinction a manager would
consult it to settle. An instrument admitting in its own report that it cannot resolve
the question it exists to inform is the substrate gap surfacing one level above
authorship.

\paragraph{Competent at the surface, failing where theory lives} The first large-scale
lifecycle study of verified AI-authored commits --- hundreds of thousands of commits
across thousands of repositories, attributed by explicit version-control metadata rather
than by classifier --- finds an asymmetry in what the assistants can repair [Liu et al.
2026]. They net-remove shallow, pattern-level debt, the code smells, while
net-introducing the defect classes that demand reasoning about program logic and
context: runtime bugs, and security issues at nearly twice the rate they fix them. A
controlled prompt study shows the same asymmetry from the input side: as the legibility
of a benign requirement degrades --- its goal clarity, completeness, and logical
consistency --- the rate at which the generated code carries vulnerabilities rises
sharply, the failure tracking the specification rather than the difficulty of the code
[Wang 2026a]. Read against Naur [1985], this is the theory-building claim coming out of
static analyzers at scale. Generation is competent exactly where competence is a matter
of surface pattern and fails exactly where it would require the theory of the system that
no one now holds --- not the generating model in any deployable sense, and not
necessarily the human who merged it. The defect survives because the understanding that
would have caught it was never built.

These are not unrelated embarrassments. They are views of one pattern: the conventional
instruments --- productivity experiments, output telemetry, authorship attribution ---
are reading a development process whose binding quantity has moved somewhere they do not
look. Even the proliferation of names for the hazard is a symptom of the same condition.
Three coinages inside a single year for substantially one quantity is the signature of a
field circling a phenomenon it has not yet pinned to an instrument --- and the
practitioners doing the naming half-see it, the same Technology Radar that foregrounds
cognitive debt reaching for Fowler's semantic diffusion to describe the churn around it
[Thoughtworks 2026; cf.\ Fowler 2006 on semantic diffusion]. Measured quantities
converge on a single name the way temperature did once the thermometer existed;
unmeasured ones breed synonyms.

\section{What survives: re-grounding on comprehension}

If authorship can no longer evidence understanding, the quantity the metrics were
proxying does not vanish with the proxy. It becomes more important and harder to see at
once. The question \emph{who can still predict, locate, and defend what this system does}
is precisely the question that comes due when an AI-built system has to be debugged,
modified, or defended --- and it is precisely the question the authorship metrics
promised to answer and can no longer answer. The quantity is Naur's [1985] theory: the
living understanding of the system held by humans the organization can actually deploy. I
will call it \emph{theory retention,} and I will say only what the collapse forces,
declining to specify the instrument that would read it.

What the collapse forces is the re-grounding. Theory retention has to be read from
evidence of comprehension rather than evidence of authorship --- from what a developer
can demonstrably do, not from what they have touched. The kinds of evidence that qualify
are familiar from the program-comprehension literature: a developer's ability to predict
the system's behavior under inputs they have not seen, to locate where a specified change
must be made, to tell the load-bearing behaviors from the incidental ones, to reconstruct
the rationale behind a design decision that does not explain itself. These are evidence of
held theory in a way that an authorship footprint, after the collapse, is not --- because
they are produced by the understanding directly, rather than by an act that used to travel
with it.

This is not virgin territory, and I want to be exact about where it stands, because the
distance between what exists and what is needed is the actual frontier. Comprehension has
been measured, with validated instruments, at the scale of the individual and the
fragment: from physiological measures of comprehension effort [Siegmund et al. 2014], to
reading-behavior metrics for the understandability of a code unit to a particular reader
[Gao et al. 2025], to controlled demonstrations that a comprehension quantity can be
degraded by generation while the conventional output signals stay green, and protected by
an intervention [Sankaranarayanan 2026], to large field studies of comprehension activity
in working professionals [Xia et al. 2018]. What does not exist --- and what the
comprehension-debt era actually requires --- is the instrument at the scale of a whole
system and a whole team, continuous and unobtrusive, because that is the only scale at
which the consequential decisions are made. The nearest public instrument that might be
mistaken for it underscores the gap rather than closing it: A.S.E [Lian et al. 2025] is a
repository-level security benchmark --- to that extent the closest public thing to a
system-scale measurement of AI-generated code --- but it scores models on whether
regenerated code reintroduces documented CVEs, a property of the generated artifact on the
security-and-surface side, not the theory a team retains. It evaluates generators, not
standing systems, and reads code, not comprehension; that it runs at repository level does
not make it the instrument required, because repository-level context is not
system-and-team scale and CVE reintroduction is not theory retention. Building it is an
open problem, and a hard one, and it is deliberately outside the scope of this paper, which
claims only that the replacement must measure comprehension and not authorship, and that
this is exactly why the existing family cannot be refined into it.

Two honest limits bound any optimism the re-grounding invites, and they should be stated
now rather than discovered later. A comprehension measure is gameable in ways a running
test is not --- comprehension can be performed without being held, the way a student
passes an examination they do not understand --- so any such instrument has to be treated
as an estimate carrying an adversarial discount, never a clean reading. And scaling a
valid comprehension measure from the fragment to the system and team, continuously, may
not admit a fully satisfactory solution; it is possible that theory retention is destined
to be estimated coarsely rather than measured precisely. The claim here survives both
limits, because it is comparative, not absolute: a coarse but honest measure of
comprehension beats a precise measurement of a substrate that has decoupled from the
target. The authorship metric's precision is now precision about the wrong thing, which is
the worse failure --- coarse and correct outranks sharp and pointed at nothing.

\section{A prediction that can be wrong}

A negative claim carrying this much weight should hand its critics a way to kill it, and
the substrate collapse makes a prediction the authorship metrics, on their own terms,
cannot. The prediction is the experiment that separates the dead substrate from the live
one.

\begin{leftbar}
\noindent\textbf{Prediction.} Across a population of AI-generated or AI-heavy systems, the
authorship-derived truck factor will fail to predict incident-resolution outcomes, while a
comprehension-grounded measure of theory retention will predict them. Concretely: systems
with a \emph{healthy} authorship truck factor --- apparently well-distributed authorship,
no obvious bus-factor exposure --- but \emph{low} comprehension-measured retention will
suffer disproportionate incident-resolution failures (longer time-to-resolution, more
escalation, more failed first fixes) when a novel incident demands theory of the system.
The authorship metric will rate these systems safe. They will not be safe. The
comprehension measure will have seen it coming, and the authorship measure will not.
\end{leftbar}

The prediction is refutable in the direction that would vindicate the metrics: if
authorship-derived truck factor goes on predicting incident-resolution outcomes on
AI-heavy systems as well as it did before the collapse, the substrate has not decoupled
and this paper is wrong. A claim earns its standing by making a prediction like this and
surviving the attempt to break it.

One caveat keeps this honest. The experiment cannot be run until the
comprehension-retention instrument of \S5 exists, and it does not yet. The prediction is
therefore a commitment the construct makes rather than a result it reports --- specified
now so that building the instrument is also building the means to refute the paper. The
unfalsifiability is in practice, not in principle: the claim is sharp enough to be killed,
and it becomes runnable the moment the instrument does.

A second, softer prediction follows from the same logic and ties back to \S4. Teams whose
system behavior grows faster than their comprehension-measured retention will show the
false-confidence signature already visible in the METR perception gap: conventional
metrics stable or improving --- tests passing, pipelines green, velocity climbing,
authorship healthily distributed --- followed by a discontinuous failure on a novel
incident none of those signals anticipated. The conventional instruments register nothing
as the exposure opens, because not one of them is measuring the quantity that is eroding.
This is the comprehension-debt literature's ``false confidence'' given a mechanical
statement rather than a cautionary one: the calm is not evidence of safety. It is evidence
that the instruments are looking elsewhere.

\section{Scope, and what this paper does not claim}

The argument is deliberately narrow, and a few disclaimers keep it honest.

This paper does not claim that AI-generated code is necessarily ill-understood --- only
that authorship can no longer tell you whether it is. A thoroughly reviewed generated
module may be understood completely; the point is that its footprint does not testify to
that, any more than an unreviewed module's footprint testifies to the reverse. The
collapse is in the inference, not in the code.

This paper does not propose a comprehension instrument, specify a probe, or describe a
mechanism for measuring theory retention at system and team scale. It asserts that such an
instrument must be grounded in comprehension, and it bounds the difficulty of building
one; it does not build it, and a reader who came for the construction will not find it
here. That absence is a choice, not an oversight.

This paper does not claim the truck factor was ever a perfect proxy, nor that it is
worthless on code still written by humans who understand it. On a pre-collapse codebase
the inference holds and the metric measures what it always did, imperfectly and
correlatedly. The claim is about the regime AI generation creates, and about the specific
error of carrying a pre-collapse instrument into it unexamined.

This paper also does not engage the fast-moving literature on making AI code generation
itself safer and more reliable: hardening generated code against vulnerabilities,
disciplining agents with reflection and verification loops, and securing the tooling and
protocols through which agents act. (Benchmarks that \emph{measure} generated code, as
against improving it, are weighed where they belong, in \S5.) That work is real and
consequential, and it is orthogonal to the argument here. Every result in it concerns the
generator --- how to make the agent produce better code, or to keep its execution within
bounds --- and none of it bears on the inference this paper turns on, because no
improvement to what a generator produces restores the link between a human's authorship
footprint and that human's understanding. However safe and self-correcting the generation
becomes, that safety is a property of the generator; to read a human's retained theory off
the generator's competence would be the very substrate error \S3 forbids. Securing the
generation is a different axis from measuring the comprehension that survives it, and the
second is this paper's only subject.

And the credited inheritances, for the record. The theory-building account is Naur's
[1985]. The metrics belong to their authors [Avelino et al. 2016; Fritz et al. 2014;
Rigby et al. 2016]. Comprehension debt as a named phenomenon belongs to Gorman [2025],
Osmani [2026], and the peer-reviewed treatment of Ahmad [2026]; the cognitive- and
epistemic-debt variants to Storey [2026], Thoughtworks [2026], and Sankaranarayanan
[2026]. The empirical figures are METR's, Faros's, and Liu et al.'s, and the
comprehension-measurement instruments are their respective authors'. What this paper
contributes is the conclusion drawn across them: that AI generation severs the authorship
inference, that the severance invalidates the knowledge-concentration family as a class
and cannot be undone by reweighting, that the field's own measurement failures are
consistent with it, and that the experiment of \S6 will bear it out.

\section{Conclusion}

For most of software's history, measuring who authored a system was a defensible way to
estimate who understood it, because authoring a thing generally meant understanding it,
and the version-control record was a fossil of the second dressed as a record of the
first. AI generation has pried the two acts apart. The fossil remains; the thing it was a
fossil of is no longer reliably there. A truck factor computed over an AI-generated
codebase is not a wrong measurement of knowledge concentration --- it is an accurate
measurement of authorship concentration, mistaken for one of knowledge concentration, in
a world where the two have quietly come apart.

The field is at the moment fever was in before the thermometer: the hazard named, named
several times over, and the exhortation begun --- review more, understand before
delegating, keep humans in the loop. None of that is wrong, and none of it is measurement.
The instrument the moment requires has to read comprehension rather than authorship, and
the first step toward building it is to stop computing the instrument that no longer reads
the right substrate, and to be clear about why. That clarity is what this paper offers,
and the whole of what it offers. The replacement is unbuilt. Naming why the old one is
dead is the precondition for building it --- and it is where the work begins.

\section*{References}
\begin{list}{}{%
  \setlength{\leftmargin}{1.6em}\setlength{\itemindent}{-1.6em}%
  \setlength{\listparindent}{0pt}\setlength{\itemsep}{0.45em}%
  \setlength{\parsep}{0pt}\setlength{\topsep}{0.4em}}
\small

\item Ahmad, M. O. (2026). Comprehension debt in GenAI-assisted software engineering
projects. \emph{30th International Conference on Evaluation and Assessment in Software
Engineering (EASE 2026)}, Glasgow. arXiv:2604.13277.

\item Avelino, G., Passos, L., Hora, A., \& Valente, M. T. (2016). A novel approach for
estimating truck factors. \emph{24th International Conference on Program Comprehension
(ICPC)}, 1--10. arXiv:1604.06766.

\item Faros AI. (2025, July). The AI Productivity Paradox: AI Coding Assistants Increase
Developer Output, But Not Company Productivity. \url{https://www.faros.ai/ai-productivity-paradox}

\item Faros AI. (2026, March). AI Engineering Report 2026: The Acceleration Whiplash.
\url{https://www.faros.ai/research/ai-acceleration-whiplash}

\item Fowler, M. (2006). Semantic diffusion. \url{martinfowler.com/bliki/SemanticDiffusion.html}.

\item Fritz, T., Murphy, G. C., Murphy-Hill, E., Ou, J., \& Hill, E. (2014).
Degree-of-knowledge: Modeling a developer's knowledge of code. \emph{ACM Transactions on
Software Engineering and Methodology}, 23(2).

\item Gao, H., Hijazi, H., Medeiros, J., Durães, J., Lam, C. T., de Carvalho, P., \&
Madeira, H. (2025). NRevisit: A cognitive behavioral metric for code understandability
assessment. \emph{EASE 2025}, Istanbul. arXiv:2504.18345.

\item GitHub. (2025). Spec-driven development with AI: Get started with a new open-source
toolkit (Spec Kit). The GitHub Blog.
\url{https://github.blog/ai-and-ml/generative-ai/spec-driven-development-with-ai-get-started-with-a-new-open-source-toolkit/}

\item Gorman, J. (2025, September 30). Comprehension debt: The ticking time bomb of
LLM-generated code. Codemanship's Blog.
\url{https://codemanship.wordpress.com/2025/09/30/comprehension-debt-the-ticking-time-bomb-of-llm-generated-code/}

\item Lian, K., Wang, B., Zhang, L., Chen, L., Wang, J., Zhao, Z., Yang, Y., et al. (2025).
A.S.E: A repository-level benchmark for evaluating security in AI-generated code.
arXiv:2508.18106.

\item Liu, Y., Widyasari, R., Zhao, Y., Irsan, I. C., \& Lo, D. (2026). Debt behind the AI
boom: A large-scale empirical study of AI-generated code in the wild. arXiv:2603.28592.

\item METR (Becker, J., Rush, N., Barnes, B., \& Rein, D.). (2025, July 10). Measuring the
impact of early-2025 AI on experienced open-source developer productivity. arXiv:2507.09089.
\url{https://metr.org/blog/2025-07-10-early-2025-ai-experienced-os-dev-study/}

\item METR (Becker, J., Rush, N., Cunningham, T., Rein, D., \& Mahamud, K.). (2026,
February 24). We are changing our developer productivity experiment design.
\url{https://metr.org/blog/2026-02-24-uplift-update/}

\item Naur, P. (1985). Programming as theory building. \emph{Microprocessing and
Microprogramming}, 15(5), 253--261.

\item Osmani, A. (2026, March 14). Comprehension debt --- the hidden cost of AI-generated
code. \url{https://addyosmani.com/blog/comprehension-debt/} (reposted, O'Reilly Radar,
April 2026)

\item Rigby, P. C., Zhu, Y. C., Donadelli, S. M., \& Mockus, A. (2016). Quantifying and
mitigating turnover-induced knowledge loss. \emph{38th International Conference on Software
Engineering (ICSE)}, 1006--1016.

\item Sankaranarayanan, S. (2026). Mitigating ``epistemic debt'' in generative
AI-scaffolded novice programming using metacognitive scripts. arXiv:2602.20206.

\item Siegmund, J., Kästner, C., Apel, S., Parnin, C., Bethmann, A., Leich, T., Saake, G.,
\& Brechmann, A. (2014). Understanding understanding source code with functional magnetic
resonance imaging. \emph{36th International Conference on Software Engineering (ICSE)},
378--389.

\item Storey, M.-A. (2026). From technical debt to cognitive and intent debt: Rethinking
software health in the age of AI. arXiv:2603.22106; also in ACM Queue.

\item Thoughtworks. (2026, April 15). Technology Radar, Volume 34.
\url{https://www.thoughtworks.com/radar}

\item Wang, B., Yu, W., Zhong, Y., Yu, H., Lian, K., Lu, C., Zheng, H., Zhang, D., \& Li,
H. (2025). AI code in the wild: Measuring security risks and ecosystem shifts of
AI-generated code in modern software. arXiv:2512.18567.

\item Wang, B., Zhong, Y., Wan, M., Yu, W., Ouyang, Y., Wu, H., Huang, Y., \& Li, H.
(2026a). Is your prompt poisoning code? Defect induction rates and security mitigation
strategies. arXiv:2510.22944.

\item Wang, B., Quan, J., Yu, X., Hu, H., Yuhao, \& Tsang, I. (2026b). Reflection-driven
control for trustworthy code agents. \emph{AAAI Conference on Artificial Intelligence
(AAAI 2026)}. arXiv:2512.21354.

\item Xia, X., Bao, L., Lo, D., Xing, Z., Hassan, A. E., \& Li, S. (2018). Measuring
program comprehension: A large-scale field study with professionals. \emph{IEEE
Transactions on Software Engineering}, 44(10), 951--976.

\end{list}

\end{document}